\documentclass[aps,showpacs,twocolumn]{revtex4}
\usepackage{graphicx}
\def\bsigma{\mbox{\boldmath $\sigma$}}

\begin{document}
\title{Resonant and non-dissipative tunneling \\
in independently contacted graphene structures}
\author{F. T. Vasko}
\email{ftvasko@yahoo.com}
\affiliation{QK Applications,  3707 El Camino Real, San Francisco, CA 94033, USA}
\date{\today}

\begin{abstract}
The tunneling current between independently contacted graphene sheets separated by boron nitride insulator is calculated. Both dissipative tunneling transitions, with momentum transfer due to disorder scattering, and non-dissipative regime of tunneling, which appears due to intersection of electron and hole branches of energy spectrum, are described. Dependencies of tunneling current on concentrations in top and bottom graphene layers, which are governed by the voltages applied through independent contacts and gates, are considered for the back- and double-gated structures. The current-voltage characteristics of the back-gated structure are in agreement with the recent experiment [Science {\bf 335}, 947 (2012)]. For the double-gated structures, the resonant dissipative tunneling causes a ten times enhancement of response which is important for transistor applications.
\end{abstract}

\pacs{72.80.Vp, 73.40.Gk, 85.30.Mn}
\maketitle

\section{Introduction}
In contrast to the tunneling processes between bulk materials, \cite{1} the tunneling  between low-dimensional systems must be assisted by scattering in order to satisfy the momentum and energy conservation laws, see results and discussions for double quantum wells or wires in Refs. 2 or 3, respectively. When the splitting energy between 2D states (this energy $\Delta$ is determined by transverse voltages applied across structure) exceeds the collision broadening energy $\Gamma$ ($\hbar /\Gamma$ is the departure time), the tunneling probability appears to be proportional $\Gamma$. In conditions of tunneling resonanse, when $\Delta\ll\Gamma$,this probability is proportional to $\hbar /\Gamma$, \cite{4} i.e. the tunnel current depends on the scattering time in the same way as the current in metallic conductor. The breakdown of the dissipative tunneling regime is possible if the energy spectrum branches are intersected and the energy-momentum conservation laws are satisfied without scattering. For example, the intersection of the parabolic electron branches in double quantum wells takes place if the magnetic field is applied perpendicular to the tunneling direction, see \cite{5} and \cite{6} for the experimental data and theory. Similar intersection between the linear branches of gapless energy spectra should take place in graphene/boron nitride/graphene (G/BN/G) heterostructure. Such a structure was reported recently \cite{7} and the tunneling transistor, which is based on the independently-contacted G-sheets connected through  a few monolayer BN, was demonstrated. \cite{8} In contrast to the semiconductor heterostructure case, the independently-contacted G/BN/G structures can be easily realized with the use of the single-layer-transfer technology. \cite{9} But a complete theoretical investigation of tunneling current in such a structure is not performed yet (some numerical results on the tunneling conductance were reported recently \cite{9a} but the current-voltage characteristics were not analyzed) and a problem is timely now.
\begin{figure}[tbp]
\begin{center}
\includegraphics[scale=0.55]{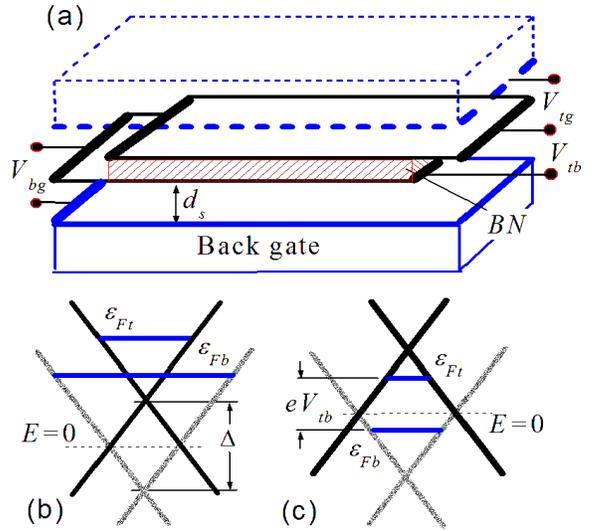}
\end{center}
\addvspace{-0.5 cm}
\caption{(Color online) (a) G$_t$/BN/G$_b$ structure under voltages $V_{tb}$ and $V_{bg}$ applied to top ($t$-) and bottom ($b$-) layers through independent contacts; back gate (blue) is separated by substrate of thickness $d_s$ and top gate under voltage $V_{tg}$ is shown by blue dashed lines. (b) Dispersion laws of $t$- and $b$-layers (black and gray crossed lines) with Fermi energies $\varepsilon_{Ft}$ and $\varepsilon_{Fb}$ for the electron-electron tunneling regime. (c) The same as in (b) for the electron-hole tunneling regime. }
\end{figure} 

In the paper, we calculate the tunneling current $I$ between the independently-contacted top (G$_t$) and bottom (G$_b$) graphene layers separated by BN. We analyze the dependencies of $I$ on the sheet concentrations (Fermi energies) and on $\Delta$, which are determined by the gate voltages applied to the contacts, see Fig. 1a. Depending on voltages applied, one can realized either  electron-electron (hole-hole) or electron-hole regimes of tunneling, as it is shown in Figs. 1b and 1c, respectively. In the latter case the cross-point $E=0$ is located between the Fermi energies ($\varepsilon_{Ft}>0>\varepsilon_{Fb}$ or vise versa) and {\it the non-dissipative regime} of tunneling takes place in addition to {\it the resonant dissipative tunneling} transitions. The current-voltage characteristics appears to be different for these regimes. For the back-gated structure, the results are in agreement with the experimental data. \cite{8} For double-gated structures, the resonant dissipative tunneling regime can be realized, with a ten times enhancement of response.

The paper is organized as follows. In Sec. II, we present the basic equations which describe the two regimes of interlayer tunneling. In Sec. III we analyze the current-voltage characteristics and compare the results for the back-gated structure with the of experimental data. \cite{8} The last section includes the discussion approximations used, and the conclusions. In Appendix, we evaluate the effective tunneling Hamiltonian for G/BN/G structure.

\section{Basic equations}
Under consideration of G$_t$/BN/G$_b$ structure, we use the tunneling Hamiltonian which connects G$_t$ and G$_b$ layers described by $2\times$2 matrices $\hat h_{Gt,b}$, i.e. we introduce $4\times$4 matrix \cite{10}
\begin{equation}
\hat H_{GBNG}  = \left| {\begin{array}{*{20}c}
   {\hat h_{Gt} } & {\hat \tau }  \\  {\hat \tau } & {\hat h_{Gb} }  
\end{array}} \right| \equiv \hat H_G  + \hat T .
\end{equation}
Here we have separated the Hamiltonian of uncoupled layers, $\hat H_G$, and the tunneling contribution, $\hat T$, written through the $2\times$2 matrix $\hat\tau =\hat\tau ^+$ which is determined by a stacking geometry of the structure, see Appendix. The charge density in G$_t$ and G$_b$ layers, $Q_t^{(k)}$ (here and below $k=t,b$), and the tunnel current density, $I_t$, are determine through the $4\times$4 density matrix $\hat\rho_t$ by the formulas \cite{11}
\begin{equation}
\left| {\begin{array}{*{20}c}
   {Q_t^{(k)} }  \\   {I_t } \end{array}} \right| =\frac{4}{L^2}{\rm Sp}\left( {\left| {\begin{array}{*{20}c}
   {e\hat P_k }  \\  {\hat I} \end{array}} \right|\hat \rho _t } \right) .
\end{equation}
Here $L^2$ is the normalization area, $\hat P_k$ is the projection operator on the $k$-states, and the interlayer current operator $\hat I$ is determined from the charge conservation requirement $I_t =dQ_t^{(t)}/dt=-dQ_t^{(b)}/dt$ (see similar calculations in Refs. 6), so that
\begin{equation}
\hat I = \frac{e}{\hbar }\left| {\begin{array}{*{20}c}
   0 & { - i\hat \tau }  \\
   {i\hat \tau } & 0  \end{array}} \right| .
\end{equation}
As a result, tunneling processes are described by the above-introduced matrices $\hat T$ and $\hat I$, as well as the density matrix governed by the standard equation: $i\hbar (\partial\hat\rho_t /\partial t) =[\hat H_{GBNG},\hat\rho_t ]$. \cite{11}

Further, we separate the diagonal and non-diagonal parts of the density matrix $\hat\rho_t \equiv [\hat\rho_t ]+\{\hat\rho_t\}$ which describe the distribution of carriers in G$_t$- and G$_b$-layers and the tunneling current, $I_t =(4/L^2 ){\rm Sp}\left(\hat{I}\{\hat \rho_t\}\right)$, respectively. Similarly to Ref. 6, one express $\{\rho\}$ through the carrier distributions determined by $[\rho ]$ and the tunneling current density takes form
\begin{equation}
I\approx\frac{4i}{L^2\hbar}\int\limits_{-\infty}^0 dt {\rm Sp}\left( [\hat\rho ]\left[ \hat T,e^{-i\hat H_G t/\hbar}\hat Ie^{i\hat H_G t/\hbar}\right] \right) .
\end{equation}
Here ${\rm Sp}(\ldots )$  means  both summations over states of carriers in G$_t$- and G$_b$-layers and averaging over lateral disorder which should be included in the Hamiltonians $\hat h_{Gt,b}$. Calculations of ${\rm Sp}(\ldots )$ are performed below with the use of the basis formed by 2-row wave functions in $k$th layer $\Psi_{\bf x}^{(k\alpha )}$ determined by the eigenvalue problem $\hat h_{Gk}\Psi_{\bf x}^{(k\alpha )}=\varepsilon_{\alpha}\Psi_{\bf x}^{(k\alpha )}$. We introduce the spectral density matrix, labeled by $l,l'=$1 and 2, as
\begin{equation}
A_{E,ll'}^{(k)}\left({\bf x}-{\bf x}'\right) =\left\langle\sum\limits_\alpha \delta \left( E -\varepsilon_{k\alpha} \right)\Psi_{l{\bf x}}^{(k\alpha )} \Psi _{l'{\bf x}'}^{(k\alpha )*} \right\rangle_k ,
\end{equation}
where the averaging over random disorder in $k$th layer $\langle\ldots\rangle_k$ is performed. Using the Fermi distribution for heavily-doped layers, when $[\rho ]$ is replaced by the $\theta$-function $\theta (\varepsilon_{Fk}-\varepsilon_{k\alpha})$ with the Fermi energies $\varepsilon_{Fk}$ in Eq.(4), we transform $I$ into 
\begin{eqnarray}
I=\frac{8\pi |e|}{\hbar L^2}\int\limits_{\varepsilon _{Fb}}^{\varepsilon_{Ft}} dE \int d{\bf x}\int d{\bf x}' \\
\times{\rm tr}\left[\hat\tau^+\hat A_E^{(b)}({\bf x}'-{\bf x})\hat \tau\hat A_E^{(t)} ({\bf x}-{\bf x}')\right] .  \nonumber
\end{eqnarray}
where ${\rm tr}(\ldots )$ means summation over the matrix variable. 

Below, we express $\hat A_E^{(k)}$ in the momentum representation as $\hat A_{E,{\bf p}}^{(k)}=i\left(\hat G_{E,{\bf p}}^{R(k)} -\hat G_{E,{\bf p}}^{R(k)~+}\right)/2\pi $, where $\hat G_{E,{\bf p}}^{R(k)}$ is the retarded Green's function of $k$th layer with the cross point energies $\pm \Delta /2$ written through the level-splitting energy. Within the model of short-range disorder 
with the same statistically independent characteristics for $k=t$ and $b$, the Green's function in the Born approximation takes form \cite{12}
\begin{eqnarray}
\hat G_{E,{\bf p}}^{R(k)} = \hat P_{\bf p}^{(+)} G_{E ,p}^{(k)}+ 
\hat P_{\bf p}^{(-)} G_{E,-p}^{(k)}\equiv{\cal G}_{Ep}^{(k)}+\frac{\hat{\bsigma}\cdot{\bf p}}{p}\overline{\cal G}_{Ep}^{(k)} , \nonumber \\
G_{E ,p}^{(k)} \approx\left[ (E\mp\Delta /2)(1 + \Lambda_{E\mp\Delta /2} +ig) -\upsilon p\right]^{-1} , ~~~
\end{eqnarray}
where $-\Delta /2$ and $+\Delta /2$ are correspondent to $t$- and $b$-layers and $\upsilon\simeq 10^8$ cm/s is the carrier velocity. The projection operators on the conduction ($+$) and valence ($-$) band states, $\hat P_{\bf p}^{(\pm )}=\left[ 1\pm (\hat{\bsigma}\cdot{\bf p})/p\right] /2$, are  written through the 2$\times$2 isospin Pauli matrix $\hat{\bsigma}$. The self-energy contributions $(E\mp\Delta /2)(\Lambda_{E\mp\Delta /2} +ig)$ are written through the logarithmically-divergent real correction which is proportional to $\Lambda_{E} =(g\pi ) \ln \left( E_c /|E|\right)$ and the coupling constant $g$. This approach corresponds to the short-range scattering with the cut-off energy $E_c$. \cite{13} The interlayer tunneling is described by the parameters
\begin{equation}
T^2 ={\rm tr}(\hat\tau^+\hat\tau ), ~~~~ T_s^2  = \frac{1}{2}\sum\limits_\mu{\rm tr}(\hat \tau ^ +  \hat \sigma _\mu  \hat \tau \hat \sigma _\mu  ) ,
\end{equation}
which appear under calculation of the matrix trace in Eq. (6). Using Eqs. (7) and (8) we transform the tunneling current density (6) into the form 
\begin{eqnarray}
I =\frac{2e}{\pi\hbar}\int\limits_{\varepsilon_{Fb}}^{\varepsilon_{Ft}}  
\frac{dE}{L^2}\sum\limits_{\bf p}\left( T^2{\rm Im}{\cal G}_{Ep}^{(t)}{\rm Im}{\cal G}_{Ep}^{(b)} \right.   \nonumber  \\
\left. +T_s^2 {\rm Im}\overline{\cal G}_{Ep}^{(t)}{\rm Im}\overline{\cal G}_{Ep}^{(b)} \right)  , ~~~~~~ 
\end{eqnarray}
where the explicit expressions for ${\cal G}_{Ep}^{(k)}$ and $\overline{\cal G}_{Ep}^{(k)}$ are determined by Eq. (7). Due to the in-plane symmetry of the problem, $I$ is written as the double-integral over the ${\bf p}$-plane and the energy interval $(\varepsilon_{Ft},\varepsilon_{Fb})$. 

Integrations in (9) are performed analytically for the collisionless case, $g\to 0$, and the tunneling current density is given by the sum of the dissipative current, which is $\propto\delta_\Gamma (\Delta ),$ and the non-dissipative contribution ($\propto\chi |\Delta |$):
\begin{eqnarray}
I\simeq J_T \left\{ {\begin{array}{*{20}c}
   {\left( {\varepsilon _{Ft}^2  - \varepsilon _{Fb}^2 } \right)\delta_\Gamma (\Delta ) ~~~~,} & {\varepsilon _{Ft} \varepsilon _{Fb}  > 0}  \\
   {{\rm sign}(\varepsilon _{Ft} )\left( {\varepsilon _{Ft}^2  + \varepsilon _{Fb}^2 } \right)} & {}  \\
   { \times \delta_\Gamma (\Delta ) + \chi |\Delta |/2 ~~~~,} & {\varepsilon _{Ft} \varepsilon _{Fb}  < 0}  \end{array}} \right.  ,   \\
J_T =\frac{{|e|\left( {T^2  + T_s^2 } \right)}}{{2\hbar ^3 \upsilon ^2 }}, ~~~~\chi =\frac{{T^2 -T_s^2 }}{{T^2 +T_s^2 }}    . \nonumber
\end{eqnarray}
Here the resonant dissipative contribution (at $\varepsilon_{Ft}\varepsilon_{Fb}>0$) is written through the $\delta$-like function $\delta_\Gamma (\Delta )=\Gamma /[\pi (\Delta^2 +\Gamma^2 )]$ with the phenomenological broadening $\Gamma\simeq g(\varepsilon_{Ft}+\varepsilon_{Fb})$. At $\varepsilon_{Ft} \varepsilon_{Fb}<0$, the non-dissipative contribution is written through the factor $\chi$ ($0<\chi <1$ because $T^2\geq T_s^2$) determined by the tunneling energies $T$ and $T_s$. These parameters depend on the stacking order of the G$_t$/BN/G$_b$ structure and rough estimate for the case of $N$-layer BN barrier \cite{14} is determined by Eq. (A.4), see Appendix. As a result, we obtain  $T,T_s\sim\gamma (\gamma /\widetilde\varepsilon_{BN})^N$, where $\gamma\sim$0.4 eV is the interlayer overlap integral and $\widetilde\varepsilon_{BN}\leq$2.5 eV is of the order of the $c$- and $v$-band energies in BN. We use below $T,T_s\sim$13 $\mu$eV for $N=4$ and $\sim$6 $\mu$eV for 6-layer BN barrier, which are in agreement with the experimental data of Ref. 8. The scattering parameters used in Eqs. (9, 10) are taken from the conductivity measurements, see similar considerations in Refs. 16. We choose the cut-off energy $E_c\sim$0.2 eV and the coupling constants $g\simeq$0.3 or 0.15 corresponded to the maximal sheet resistance $\sim$4 or 2 k$\Omega$ per square.
\begin{figure}[tbp]
\begin{center}
\includegraphics[scale=0.55]{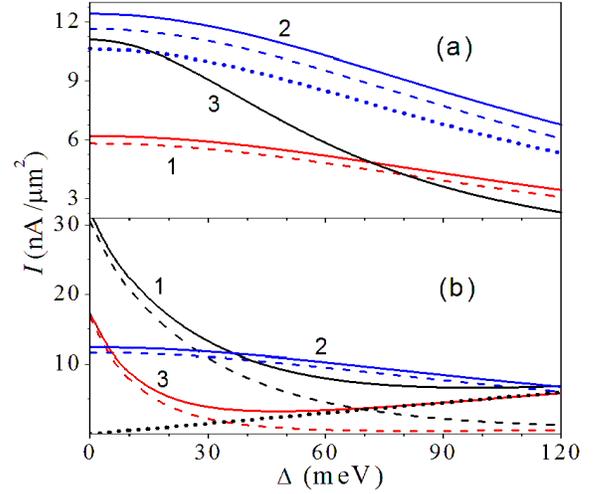}
\end{center}
\addvspace{-0.5 cm}
\caption{(Color online) (a) Tunneling current density $I$ versus level splitting $\Delta$ for the resonant tunneling regime at $\varepsilon_F =$200 meV and $\delta\varepsilon =$50 meV (1) or 100 meV (2) for $g=0.3$. Solid and dashed curves are plotted for $\chi =$1 and $\chi =$0, respectively. Curve (3) is plotted for parameters (1) at $g=0.15$. Dotted curve fits dependencies (2) with the use Eq. (10).  (b) The same as in panel (a) for the non-dissipative regime, if $\varepsilon_F =$0 and $\delta\varepsilon =$200 meV (1) or 100 meV (3). Curve (2) corresponds  $\varepsilon_F =$200 meV, $\delta\varepsilon =$100 meV and dotted line gives $\propto\Delta$ contribution in Eq. (10).  }
\end{figure}  

The current density $I$ is dependent on $\varepsilon _{Ft}$, $\varepsilon _{Fb}$ and $\Delta$, which are determined by the drops of voltages applied to three- or four-terminal structures, see Fig. 1a. Before study of the current-voltage characteristics, we consider the dependencies of the total current $I$ on these three energies. In Fig. 2 we plot the tunneling current, which is determined by Eqs. (9) and (7), versus $\Delta$ for different Fermi energies determined by $\varepsilon_{F}=(\varepsilon_{Ft}+\varepsilon_{Fb})/2$ and $\delta\varepsilon_{F}=\varepsilon_{Ft}-\varepsilon_{Fb}$ at $\chi =$1 or 0. The resonant dissipative regime of tunneling is realized at $|\varepsilon_{F}|>|\delta\varepsilon_{F}/2|$ and peak value $I(\Delta =0)$ increases both with doping levels and with $g$, as it is shown in Fig. 2a. If $|\varepsilon_{F}|<|\delta\varepsilon_{F}/2|$, the electron-hole tunneling regime takes place and the non-dissipative contribution becomes dominant with increasing of $\Delta$, where $I\propto\chi |\Delta |$. At small $\Delta$ the dependency $I(\Delta )$ is transformed into narrow peak due to the dissipative contribution with weak broadening, see Eq. (10) and Fig. 2b. The numerical results given by Eq. (9) are in a good agreement with the approximations (10), shown by the dotted asymptotics, because of weak ($\leq$10\%) contributions from the renormalization of energy spectra. 

Neglecting the quantum capacitance contributions and the near-contact drops of potentials, we use $\varepsilon_{Ft}-\varepsilon_{Fb}\simeq eV_{tb}$ and the Gauss theorem connected the carrier concentrations in graphene with interlayer electric fields, $V_{tb}/d$ and $V_{bg}/d_s$ (here $d$ and $d_s$ are thicknesses of BN layer and SiO$_2$ substrate). As a result, the Fermi energies and $\Delta$ are connected with drops of voltages as follows:
\begin{eqnarray}
2\varepsilon_{Ft,b}\simeq \pm eV_{tb} - F(eV_{tb} \varepsilon _d ) - F(eV_{bg} \varepsilon _{d_s }  - eV_{tb} \varepsilon _d ) ,  \nonumber \\
\Delta\simeq eV_{tb}+ F(eV_{tb}\varepsilon_d )-F(eV_{bg}\varepsilon _{d_s }-eV_{tb} \varepsilon _d ) , ~~ \nonumber  \\
F(x) ={\rm sign}(x)\sqrt{|x|} , ~~~~ \varepsilon_d =\frac{{\epsilon (\hbar v)^2 }}{{4e^2 d}} . ~~~~
\end{eqnarray}
Here ''+'' and ''$-$'' stand for G$_t$- and G$_b$-layers, $\varepsilon_{Ft,b}>0$ or $<0$ correspond to electron or hole doping, and the dielectric permittivity $\epsilon\simeq$4 is the same for BN and SiO$_2$ layers, see Refs. 7 and 8. For the double-gated structure with top gate separated by BN  insulator of thickness $d_t$, one obtains similar expressions for $\varepsilon_{Ft,b}$ and $\Delta$ after the replacement $eV_{tb}\varepsilon_d\to eV_{tb}\varepsilon_d -eV_{tg} \varepsilon_{d_t}$, see voltages shown in Fig. 1a. The approach (11) is valid for the heavily-doped graphene layers, so that the fields $V_{tb}/d$, $V_{bg}/d_s$, and $V_{tg}/d_t$ should be strong enough. On the other hand, these fields are restricted by the breakdown condition for BN layer, when these fields should be $\ll 7$ MV/cm. \cite{16} 
\begin{figure}[tbp]
\begin{center}
\includegraphics[scale=0.55]{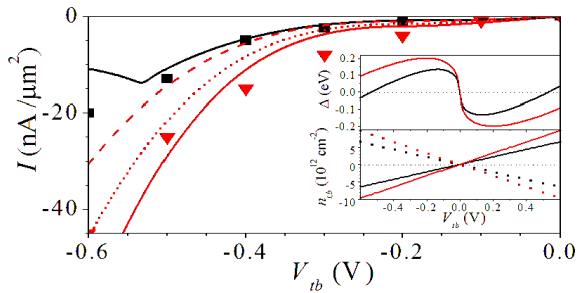}
\end{center}
\addvspace{-0.5 cm}
\caption{(Color online) Tunneling current $I$ versus $V_{tb}$ at $V_{bg}=$0 for structures with 4- and 6-layer BN barriers (black and red curves, respectively; squares and triangles are experimental points \cite{8}). Solid, dotted, and dashed curves correspond to $\chi =$1. 0.5, and 0. Insets show $\Delta$ and concentrations $n_t$ or $n_b$ (solid or dashed lines)  versus $V_{tb}$.  }
\end{figure} 

\section{Current-voltage characteristics}
In this section we analyze the current-voltage characteristics for the back- and double-gated structures. Below, the tunneling current density is determined by Eqs. (7) and (9) with the Fermi energies and the level splitting written through the voltages applied according to Eqs. (11). The parameters described both the elastic scattering processes and the interlayer tunnel coupling  are chosen the same as for the above calculations shown in Fig. 2.

For the back-gated structure with $N$-layer BN barriers ($N=$4 and 6) and $g=$0.3 the dependencies of $I$ versus $V_{tb}$ at fixed $V_{bg}$ are shown in Fig. 3 at $V_{bg}=0$, when $I_{-V_{tb}}=-I_{V_{tb}}$. \cite{17} The electron (hole) concentrations are $\propto V_{tb}$ and $\Delta$ is the sum of linear and square-root functions, see insets. The current-voltage characteristics $I(V_{tb})$ are in reasonable agreement with the experimental data of Ref. 8 if we used the scale factor $J_T$ with the above-estimated $T$ and $T_s$. The dependencies on $\chi$ are weak enough ($\leq$25\% for $N=$4 and $\leq$10\% for $N=$6, not shown). If $\Delta\sim$0, when $V_{tb}\simeq$-0.55 V for 6-layer barrier, a negative resonant contribution due to dissipative tunneling gives $\sim$30\% variations of the $I$-$V$ characteristic. Such a peculiarity was not found in Ref. 8; probably, it is due to a lateral redistributions of charges or a large-scale inhomogeneities of the samples used. The $I$-$V$ characteristics of back-gated structures at $V_{bg}\neq 0$ are plotted in Fig. 4 for $N=$4, 6 and $\chi =1$, 0. Once again, there is a reasonable agreement of $I(V_{tb},V_{bg})$ with the available experimental data for $N=$4 at $V_{tb}<0$. A visible asymmetry of $I$ takes place if $V_{tb}\to -V_{tb}$ but deviations from experimental data increase. The dependencies for $N=6$ are similar but $I$ is $\sim$2-times weaker and the resonant dissipative contribution is absent for $V_{bg}=-$25 V because $\Delta\neq 0$, see inset.
\begin{figure}[tbp]
\begin{center}
\includegraphics[scale=0.55]{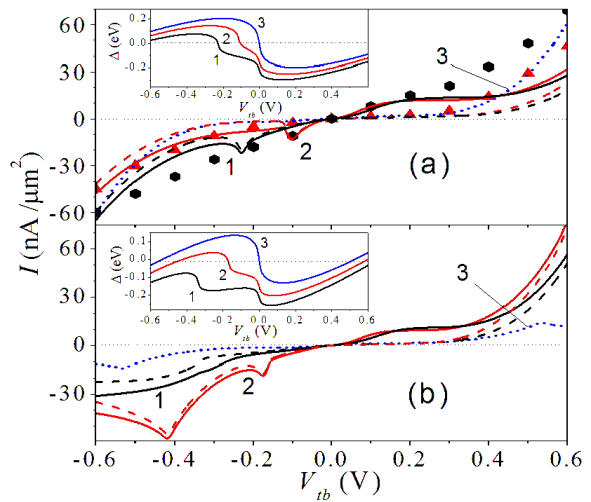}
\end{center}
\addvspace{-0.5 cm}
\caption{(Color online) (a) Current density $I$ versus $V_{tb}$ for 4-layer BN barrier at $V_{bg}=$0 (red), -25 V (blue), and -50 V (black). Experimental points are shown for $V_{bg}=$0 (triangles) and -50 V (hexagons). Inset shows $\Delta$ versus $V_{tb}$. (b) The same as in panel (a) for 6-layer BN barrier.  }
\end{figure} 

An enhancement of the resonant dissipative tunneling contributions takes place in the double-gated (four-terminal) structure, when $I$ depends on $V_{tb}$, $V_{tg}$, and $V_{bg}$. These dependencies are shown in Fig. 5 for the structure with 10-layer BN cover layer ($d_t =$3.4 nm; with a negligible tunneling, if $d_t\gg$2 nm \cite{18}) at different top- and back-gate voltages ($V_{tg}=0$ corresponds to the three-terminal structure). Since the condition $\Delta =0$ is realized now at higher energies, the resonant dissipative tunneling peaks are $>$10 times greater than the background current. This is the central result which should be important for the transistor applications. Note, that for $\Delta\sim$0 at $V_{tb}\sim$0 (the curve 3 in Fig. 5a) the resonant dissipative peak is suppressed. Beyond of this narrow region, the resonant condition $\Delta (V_{tb})\approx 0$ should not be suppressed by lateral inhomogeneities and the peaks caused by the resonant dissipative tunneling should not be smeared. 
\begin{figure}[tbp]
\begin{center}
\includegraphics[scale=0.55]{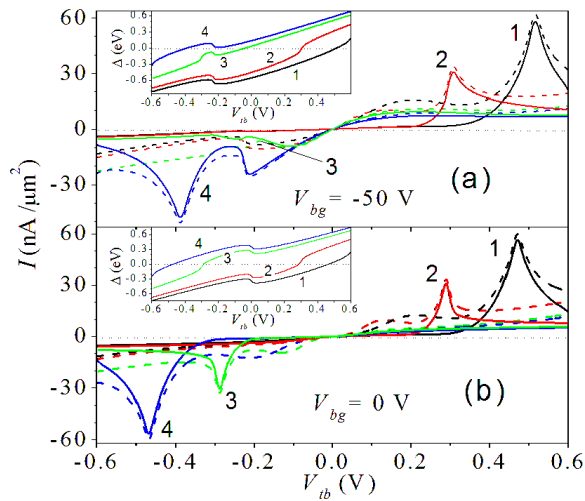}
\end{center}
\addvspace{-0.5 cm}
\caption{(Color online) (a) Dependencies $I(V_{tb})$ for double-gated structure with 4-layer BN barrier and 10-layer BN cover layer under $V_{bg}=-50$ V and $V_{tg}=1.5$ V (1), 0.75 V (2), -0.75 V (3), and -1.5 V (4). Inset shows $\Delta$ versus $V_{tb}$. Solid and dashed curves correspond to $\chi =$0 and 1, respectively. (b) The same as in upper panel for $V_{bg}=0$ V.  }
\end{figure} 

Overall, the consideration presented here gives an adequate theoretical description of the tunneling transistor effect which is in reasonable agreement with the experimental data. \cite{8} The analysis performed opens a way for the verification of scattering mechanisms and tunneling parameters in G/BN/G structures. In addition, the double-gated (four-terminal) structure was not considered before and this structure show a great (ten times) enhancement of tunneling current tunability.

\section{Conclusion}
We have adopted the theory of low-dimensional tunneling \cite{4,6} to the case of the tunneling charge transfer in independently contacted graphene structures with multi-layer BN barrier taking into account the two basic mechanisms of charge transfer: resonant dissipative tunneling and non-dissipative tunneling. Both the non-dissipative tunneling current and the non-resonant dissipative tunneling processes are responsible for the current-voltage characteristics of back-gated structures. \cite{8} The resonant dissipative tunneling regime is achieved for the double-gated structures and these devices demonstrate a ten times enhancement of $I$-$V$ characteristics which is important for transistor applications.

Further, we discuss the assumptions used. Because of a luck of data on stacking order in G/BN/G structures the tunneling energies $T$ and $T_s$ were estimated from the current-voltage characteristics of Ref. 8 and this result is in agreement with the tight-binding model. \cite{14} More accurate estimates for $T$ and $T_s$ should be based on additional structural measurements. The simplified electrostatics description, given by Eq. (11), fails for the low-doped G$_t$- or G$_b$-layers, under weak interlayer fields applied. In these narrow regions, a more complicated description, which involves the quantum capacitance effect and the contact phenomena, should be applied. The rest of assumptions (the model of elastic scattering, \cite{15} identical scattering in G$_{t,b}$-layers, weakness of long-range disorder, and the single-particle approach) are rather standard. 

To conclude, we believe that the description of tunneling processes is an essential part of physics of graphene and the results obtained can be applied for characterization of scattering mechanisms and tunneling parameters in the tunnel-coupled graphene structures. More importantly, that these results open a way for improvement of tunneling transistor, a new type of graphene-based device. We believe that our study will stimulate a further investigation of these device applications.

\appendix* 
\section{Tunneling Hamiltonian}
Below we describe the tunnel-coupled states in G/BN/G structure using the 6-column wave function $[\psi_t , \phi , \psi_b ]$ which is written through the spinors $\psi_{t,b}$ correspond to the G$_t$ and G$_b$ graphene layers. These layers are connected through the spinor $\phi$ described the single BN-layer. Within the tight-binding approach, \cite{14} the eigenstate of energy $E$ is determined by the problem written through the $6\times 6$ Hamiltonian
\begin{equation}
\left| {\begin{array}{*{20}c}
   {\hat h_{G_t}-E} & {\hat h_{GBN} } & 0  \\
   {\hat h_{GBN}^ +  } & {\hat h_{BN}  - E} & {\hat h_{BNG} }  \\
   0 & {\hat h_{BNG}^ +  } & {\hat h_{G_b}-E}  \\
\end{array}} \right|\left| {\begin{array}{*{20}c}
 \psi_t  \\  \phi  \\   \psi_b  \end{array}} \right| = 0 ,
\end{equation}
where $\hat{h}_{G_t}$ and $\hat{h}_{G_b}$ are the $2\times 2$ Hamiltonians of 
G$_t$, G$_b$, and BN-layers, while $\hat{h}_{GBN}$ and $\hat{h}_{BNG}$ describe weak interlayer coupling of G$_t$ and G$_b$ sheets with BN layer. Under a transverse voltage applied, $\hat h_{G_{t,b}}=\hat h_{G}\pm\Delta /2$, where $\hat h_{G}$ is the Hamiltonian of single graphene layer and $\Delta$ is a splitting energy between the cross-points in G$_t$- and G$_b$-layers. For the low-energy ($|E|<$1 eV) states, if $|E|\ll |\varepsilon_{c,v}|$ where $\varepsilon_c\sim$3.4 eV and $\varepsilon_v\sim$-1.4 eV are the $c$- and $v$-band extrema energies in BN, we can eliminate the spinor $\phi$ from the system (A.1). As a result, the eigenstate problem is determined by the $4\times 4$ effective tunneling Hamiltonian
\begin{equation}
\left| {\begin{array}{*{20}c}
   {\hat h_G  + \hat \tau _t  - E} & {\hat \tau }  \\
   {\hat \tau ^ +  } & {\hat h_G  + \hat \tau _b  - E}  \\
\end{array}} \right|\left| {\begin{array}{*{20}c}
\psi_t  \\  \psi_b \end{array}} \right| = 0 .
\end{equation}
Here the $2\times 2$ matrix $\hat{\tau}$ describes tunneling through BN insulator, while $\hat\tau_t$ and $\hat\tau_b$ correspond to the tunneling  renormalization of $t$- and $b$-states:
\begin{eqnarray}
\hat{\tau}\approx -\hat h_{GBN}\hat h_{BN}^{-1}\hat h_{BNG} , ~~~~~~~~~~ \\
\hat\tau_t\approx -\hat h_{GBN}\hat h_{BN}^{-1}\hat h_{GBN}^+ , ~~
\hat\tau_b\approx -\hat h_{BNG}^ + \hat h_{BN}^{-1} \hat h_{BNG} . \nonumber
\end{eqnarray}
Thus, we arrive to the Hamiltonian (1) with the renormalization contributions $\tau_{t,b}$ included to $\hat h_{G_{t,b}}$; these corrections are negligible for the weak tunneling regime.

For the case of $N$-layer BN insulator, we consider the $2(N+2)$-column wave function with $N$-spinors $\phi_1 ,\ldots ,\phi_N$ described the BN layers. These states are coupled by the interlayer hopping matrices $\hat h_{BNBN}$ and $\hat h_{BNBN}^+$ which are placed to upper and lower sub-diagonals of $2(N+2)\times 2(N+2)$ Hamiltonian. \cite{14} After eliminations of the spinors $\phi_1 ,\ldots ,\phi_N$ from the tight-binding eigenstate problem, we arrive to Eq. (A.3) where the non-diagonal matrix $\hat{\tau}$ is replaced by
\begin{equation}
\hat{\tau}_N\approx -\hat h_{GBN}(\hat h_{BN}^{-1}\hat h_{BNBN})^{N-1}\hat h_{BN}^{-1}\hat h_{BNG} .
\end{equation}
For the numerical estimates of $T$ and $T_s$ given by Eq. (8) we assume that the diagonal matrix $\hat h_{BN}^{-1}$ is of the order of $\widetilde \varepsilon_{BN}^{-1}$ where $\widetilde \varepsilon_{BN}$ is determined by $\varepsilon_{c,v}$. The hopping matrices $\hat h_{GBN}$, $\hat h_{BNG}$, and $\hat h_{BNBN}$ are estimated by the interlayer overlap integral $\gamma$ and they are strongly dependent on a stacking order of G/BN/G structure.


\begin{thebibliography}{42}
\bibitem{13}
We use the model of disorder with random potentials $V_{\bf x}$ described by the Gaussian correlator $\langle V_{\bf x}V_{\bf x'}\rangle$, with the averaged amplitude $\overline V$, and the correlation length $l_c$, if the cut-off energy $E_c\simeq\upsilon\hbar /l_c$ exceeds the energy scale under consideration. For the low-energy region, $|E|\ll E_c$, the model gives $\propto E$ relaxation rate, i. e. we deal with a short-range scattering and the coupling constant is given by $g=(\overline V^2 l_c/\hbar\upsilon )^2\pi/2$.
\bibitem{17}
From Eq. (9) one obtains $I_{-\Delta}=-I_{\Delta}$, $I_{-\delta\varepsilon_F}=-I_{\delta\varepsilon_F}$, and $I_{-\varepsilon_F}=I_{\varepsilon_F}$. According to Eq. (11), $\Delta$ and $\varepsilon_{Ft,b}$ are even functions of voltages, so that $I$ remains the same under $V_{tb, bg}\to -V_{tb, bg}$ or $V_{tg, tb, bg}\to -V_{tg, tb, bg}$.
\end{thebibliography}
\end{document}